\documentclass[a4paper,fleqn,usenatbib]{mnras}

\usepackage[T1]{fontenc}
\usepackage{ae,aecompl}

\usepackage{amsmath,amssymb,bm,upgreek,color}
\usepackage{mathrsfs}
\usepackage[latin1]{inputenc}
\usepackage{upgreek}
\usepackage{enumerate}
\usepackage{wrapfig}
\usepackage{multicol}
\usepackage{lscape}
\usepackage{pifont}
\usepackage{subfigure}
\usepackage{graphicx}
\usepackage{natbib}
\usepackage{float}
\title[Reconstruction of $f(R)$ gravity models for an accelerated
universe using Raychaudhuri equation]
{Reconstruction of $f(R)$ gravity models for an accelerated universe using Raychaudhuri equation}

\author[Shibendu Gupta Choudhury, Ananda Dasgupta, Narayan Banerjee]{
Shibendu Gupta Choudhury$^{1}$\thanks{E-mail: sgc14ip003@iiserkol.ac.in},
Ananda Dasgupta$^{1}$\thanks{E-mail: adg@iiserkol.ac.in},
Narayan Banerjee$^{1}$\thanks{E-mail: narayan@iiserkol.ac.in}\\
$^{1}$Department Of Physical Sciences, IISER Kolkata, Mohanpur, Nadia 741235, India\\
}

\date{Accepted ??. Received ??; in original form ??}
\pubyear{????}

\begin{document}

\pagerange{\pageref{firstpage}--\pageref{lastpage}} 

\maketitle

\label{firstpage}

\newcommand{\be}{\begin{equation}}
\newcommand{\ee}{\end{equation}}
\newcommand{\bea}{\begin{eqnarray}}
\newcommand{\eea}{\end{eqnarray}}

\begin{abstract}
A new strategy for the reconstruction of $f(R)$ gravity models have been attempted using Raychaudhuri equation. 
Two examples, one for an eternally accelerating universe and the other for one that mimics a $\Lambda$CDM expansion 
history have been worked out. { For both the cases, the relevant $f = f(R)$ could be found out analytically. In the first case, $f(R)$ is 
found to be a combination of power-law terms  and in the expression for the  second case involves hypergeometric functions.
The evolution history of the universe, given as specific values of the kinematical quantities like the jerk or the 
deceleration parameter, serve as the input. It is found that the corresponding $f(R)$ gravity models, in both the examples, 
are not viable options.}
\end{abstract}

\begin{keywords}
 dark energy, cosmological parameters 
\end{keywords}

\section{\label{sec:level1}Introduction}

 Arguably, the most talked about but unresolved puzzle in cosmology for the last twenty years has been that of ``Dark Energy'', 
 the driver of the alleged accelerated expansion of the universe \citep{rubin, haridasu}. Although a 
 non-zero cosmological constant can indeed match the observational data \citep{paddy}, its observationally required value 
 appears to be too  small compared to the theoretically predicted value. Quintessence models, 
 which are scalar fields with a potential, also do very well in explaining the cosmological data with a bit of fine 
 tuning, but there is hardly any strongly motivated scalar field model with support for that from 
 theoretical particle  physics. For a recent account of various dark energy models, we refer to the work of \citet{brax}. \\

A parallel approach towards finding a resolution of accelerated expansion is to modify the theory of gravity rather 
than to introduce an exotic matter. Examples of such attempts include a non-minimally coupled scalar field
theory \citep{martin, nb} or an $f(R)$ theory of gravity \citep{capozz, odinta, odintb, carroll, nb1}. In an $f(R)$ gravity model, 
the Ricci scalar $R$ in the Einstein-Hilbert action is generalised to $f(R)$, an analytic function
of $R$. It has been known that higher powers of $R$ in the Einstein-Hilbert action can give rise to ``inflation'', 
an accelerated expansion in the very early stages \citep{staro}. It is then an obvious avenue to check
if negative powers of $R$ in the action can give rise to a late time acceleration. Every single form of $f(R)$ gives 
rise to a new theory of gravity, so it is essential that the theory is tested against observations,
not only cosmological, but also other requirements such as the stability of the solutions, local astronomy like 
perihelion shift or the amount of light bending. Some  investigations along these lines are there in the
current literature, such as those in \citet{dolgov, cemb, nojiri, nojiri2}. For a comprehensive review of $f(R)$ 
gravity models, we refer to the work of \citet{faraoni}. \\

There are plenty of models that indeed fit the bill for an accelerated expansion of the universe, but there are hardly any 
that have a pressing requirement imposed by other branches of physics. In the absence of a
theoretical model that is a clear winner as  dark energy, a reconstruction of models from observational data
becomes a very good option. The idea is to find the required matter distribution from a given evolution
history of the universe \citep{ellis}. \\

In the present work, we make an attempt to reconstruct $f(R)$ gravity models, not from the observational data, but rather 
following the work of \citet{ellis}. We choose a particular form of evolution
leading to an accelerated expansion, implemented through a kinematical quantity and seek for the relevant $f(R)$ 
gravity models. For the importance of the kinematical quantities in the game of reconstruction of accelerated models,
we refer to \citet{visser, zhang, ankan}. \\

While attempts to find the relevant $f(R)$ model through this kind of reverse engineering can already be found in the 
literature \citep{Song, Pogo, capo1, Nojiri3, Dunsby, carloni, Lomb, He}, we shall adopt a different 
strategy. We utilise the Raychaudhuri equation \citep{akr, ehlers, ehlers1}, duly modified for $f(R)$ gravity, for the purpose of the 
reconstruction. We construct the kinematical quantities from the given expansion history
and write the metric components (which for a spatially homogeneous and isotropic expansion is contained in the scale 
factor only) in terms of the Ricci scalar $R$ and integrate Raychaudhuri equation for $f(R)$. \\

{Raychaudhuri equation only assumes Riemannian geometry at the outset, and thus can work equally well in $f(R)$ gravity theories. This equation can thus be very useful in  extracting some general results about the model even without actually solving for the metric. For instance, one can look at the fate of the effecive energy condition. The present case, however, is simple where even without using Raychaudhuri equation one can arrive at the results with a few more steps. For a more involved situation, this technique may lead to informations that cannot be obained otherwise. The motivation for using Raychaudhuri equation in this case is to start from a situation as general as possible.} \\ 

Raychaudhuri equation has already been utilised in the context of $f(R)$ gravity in order to look at the effective energy 
conditions and hence to assess the possibility of obtaining a repulsive gravity out of geometry
itself \citep{santos, albareti, santos2}. We employ this powerful tool of Raychaudhuri equation directly to find
the $f(R)$ gravity model for two cases. The first one is the case of an eternally accelerated model. 
The second one is the case where the evolution mimics that due to a $\Lambda$CDM model, which apparently is the most favoured
behaviour of the evolution in terms with the observational data. In the first case, we 
obtain a combination of powers of the Ricci scalar $R$. In the second case, combinations of hypergeometric functions are 
obtained. { The viability of the $f(R)$ models against various cosmological and astronomical requirements are also analysed.} \\

The paper is organised as follows. Section 2 deals with the relevant  equations in $f(R)$ gravity. In the next section 
we present the Raychaudhuri equation for a general $f(R)$ gravity model. The fourth section 
presents the actual reconstruction of $f(R)$ models for two examples, an ever accelerating universe and a $\Lambda$CDM model and 
also the viability analysis. The fifth and final section includes some concluding remarks.

\section{\label{sec:level2}$\lowercase{f}(R)$ Gravity}
The action that defines an $f(R)$ gravity theory is given by,
\begin{equation}
 S=\int \mathrm{d}^4x \sqrt{-g}f(R)+S_m,
\end{equation}
where $f(R)$ is an analytic function of the Ricci scalar $R$ and $S_m$ is the action for the relevant matter distribution. A 
variation of the action with respect to the metric tensor $g_{\alpha\beta}$ gives the following field equations,
\begin{equation}\label{fieldeq}
 f^{\prime}R_{\mu\nu} -\frac{f}{2} g_{\mu\nu}-(\nabla_\mu\nabla_\nu-g_{\mu\nu}\square)f^\prime=T_{\mu\nu},
\end{equation}
where $f^\prime(R)=\dfrac{\mathrm{d}f(R)}{\mathrm{d}R}$ and $T_{\mu\nu}=-\dfrac{2}{\sqrt{-g}}\dfrac{\delta S_m}{\delta g^{\mu\nu}}$ 
is the 
stress-energy tensor. We have chosen the units such that $c=8\pi G=1$. \\

The field equations in $f(R)$ gravity can be written in terms of the Einstein tensor with an effective 
energy-momentum tensor $T_{\mu\nu}^{\text{eff}}$, that takes care of the contribution from the curvature \citep*{Guarnizo}, as,
\begin{equation}\label{fe1}
 G_{\mu\nu}=\frac{1}{f^\prime(R)}(T_{\mu\nu}+T_{\mu\nu}^{\text{eff}}),
\end{equation}
where
\begin{equation}
 T_{\mu\nu}^{\text{eff}}=\left[\frac{f-Rf^\prime}{2}g_{\mu\nu}+(\nabla_\nu \nabla_\nu-g_{\mu\nu}\square)f^\prime\right].
\end{equation}
This equation looks like Einstein's equations, at least formally, with a difference that the presence of $f^\prime$ 
indicates a non-minimal coupling. The effective gravitational coupling will not be a constant in this formulation.

\section{Raychaudhuri equation and $\lowercase{f}(R)$ gravity}\label{Raychaudhuri}
Raychaudhuri equation for a timelike congruence having velocity vector $u^\mu$ is given by \citep{akr, ehlers, ehlers1},
\begin{equation}
\label{raych-eq}
 \frac{\mathrm{d}\theta}{\mathrm{d}\tau}=-\frac{1}{3}\theta^2+\nabla_{\mu}a^{\mu}-\sigma_{\mu\nu}\sigma^{\mu\nu}+\omega_{\mu\nu}\omega^{\mu\nu}-R_{\mu\nu}u^\mu u^\nu .
\end{equation}
where $\theta=\nabla_\mu u^\mu$ is the expansion scalar, $\tau$ is affine parameter, 
$\sigma_{\mu\nu}=\nabla_{(\nu}u_{\mu)}-\frac{1}{3}h_{\mu\nu}\theta+a_{(\nu}u_{\mu)}$
is the shear tensor where $h_{ab}$ is the spatial metric, $\omega_{\mu\nu}=\nabla_{[\nu}u_{\mu]}-a_{[\nu}u_{\mu]}$ 
is the rotation tensor, 
$a^\mu=u^\nu\nabla_{\nu}u^{\mu}$ is the acceleration vector, $R_{\mu\nu}$ is the Ricci scalar and 
$u_{\mu}$ is the timelike velocity vector. \\

Using the field equations \eqref{fieldeq} for $f(R)$ theory,  the last term in the right hand 
side of equation (\ref{raych-eq}) can be written as,
\begin{equation}
 R_{\mu\nu}u^\mu u^\nu=\frac{1}{f^\prime}\left[T_{\mu\nu}+\frac{f}{2} g_{\mu\nu}+(\nabla_\mu\nabla_\nu-g_{\mu\nu}\square)f^\prime\right]u^\mu u^\nu.
\end{equation}

The present work deals with a spatially isotropic and homogeneous universe with a flat spatial section given by the metric
\begin{equation}
 \mathrm{d}s^2 = -\mathrm{d}t^2 + a^2 (t) (\mathrm{d}r^2 + r^2 \mathrm{d}{\theta}^2 + r^2 \sin^2 \theta \mathrm{d}{\phi}^2),
\end{equation}
where $a(t)$ is the scale factor. \\

For such a metric and a matter distribution of a perfect fluid given by $T^{\mu\nu} = (\rho + p)u^{\mu}u^{\nu} + p g^{\mu\nu}$, 
Raychaudhuri equation (\ref{raych-eq}) takes the form \citep{Guarnizo}
\begin{equation}\label{rceq1}
 \frac{\ddot{a}}{a}=\frac{1}{f^\prime}\left(\frac{f}{6}+H f^{\prime\prime}\dot{R}-\frac{\rho}{3}\right).
 \end{equation}
It should be noted that we have not assumed any equation of state for the fluid distribution until now, 
but field equations (\ref{fe1}) have been used in Raychaudhuri equation (\ref{raych-eq}) so as to eliminate the fluid pressure $p$.

\section{Reconstruction of $\lowercase{f}(R)$ gravity models}\label{recon}

We shall now try to reconstruct $f(R)$ gravity models for a given mode of acceleration of the universe using 
equation (\ref{rceq1}).
The mode of acceleration will be determined by the kinematical quantities like the deceleration parameter $q$ or the 
jerk parameter $j$.
Two examples are considered here, one in which the universe is ever accelerating with a constant deceleration parameter 
and the other which 
has the jerk parameter $j=1$ indicating a model that mimics the behaviour of the  $\Lambda$CDM model in standard 
general relativity.

\subsection{A constant deceleration parameter}\label{decpa}
The Hubble parameter $H=\dfrac{\dot{a}}{a}$ is the oldest observable quantity in physical cosmology. 
As it was found to be evolving, the next higher order derivative of $a$, expressed as the deceleration parameter
$q=-\dfrac{a\ddot{a}}{{\dot{a}}^2}$, 
used to be a focus of interest.

In the present section we consider a constant negative deceleration parameter $q$, given by, 
\begin{equation}\label{q}
 q=-\frac{1}{H^2}\frac{\ddot a}{a}=\mbox{constant}=-m,
\end{equation}
where $m$ is a positive constant restricted as $0<m<1$. Equation (\ref{q}) can be integrated twice to yield a simple 
power-law solution for the scale factor as,
\begin{equation}\label{sc-fact}
 a(t)=C (t-t_0)^{\frac{1}{1-m}},
\end{equation}
where $C$ and $t_0$ are integration constants. \\

{ This model obviously describes a universe that is  ever-accelerating.  We can calculate the effective equation of state 
for this kind model using the 
equations \citep{faraoni},
\begin{equation}\label{hubble}
 H^2=\frac{\rho_\mathrm{eff}}{3}
\end{equation}
and 
\begin{equation}\label{dec}
 \frac{\ddot{a}}{a}=-\frac{3 p_\mathrm{eff}+\rho_\mathrm{eff}}{6}
\end{equation}
where $\rho_\mathrm{eff}$ and $p_\mathrm{eff}$ are the effective energy density and effective pressure respectively.
Using the solution for the scale factor \eqref{sc-fact} we get, the equation of state parameter
\begin{equation}
 w_\mathrm{eff} = \frac{p_\mathrm{eff}}{\rho_\mathrm{eff}}=-\frac{1+2m}{3}.
\end{equation}
For the two extreme values of $m$, namely $0$ and $1$, $w_\mathrm{eff}$ takes the values $-1/3$ and $-1$ respectively.

We can also look at the effective energy condition in this context if we calculate the quantity $\rho_\mathrm{eff}+3p_\mathrm{eff}$ which for this case is given by,
\begin{equation}
 \rho_\mathrm{eff}+3p_\mathrm{eff}=\frac{-6m}{(1-m)^2(t-t_0)^2}
\end{equation}
This is always negative and thus violates the energy condition, which is expected for an eternally accelerated model.
\\}

Using the solution for the scale factor \eqref{sc-fact}, equation \eqref{rceq1} can be written in terms of the scale factor $a$ as,
\begin{equation}\label{Rceq2}
\begin{split}
\frac{12(1+m)}{(1-m)^3}{\left(\frac{C}{a}\right)}^{4(1-m)}f^{\prime\prime}(R)+\frac{m}{(1-m)^2}\left(\frac{C}{a}\right)^{2(1-m)}f^{\prime}(R)
\\-\frac{f(R)}{6}=-\frac{\rho}{3}.
\end{split}
\end{equation}
For a spatially flat FRW metric, the Ricci scalar $R$ is given by  $R=6\dfrac{\dot{a}^2}{a^2}+6\dfrac{\ddot{a}}{a}.$ 
Thus, using the solution for the scale factor (\ref{sc-fact}), one can write,
\begin{equation}
R=6\frac{\dot{a}^2}{a^2}+6\frac{\ddot{a}}{a}=6\frac{(1+m)}{(1-m)^2}\left(\frac{C}{a}\right)^{2(1-m)}.
\end{equation}
Equation \eqref{Rceq2} can now be written by replacing the terms involving $a$ by powers of $R$ as,
\begin{equation}\label{Rceq3}
\frac{1-m}{3(1+m)} R^2 f^{\prime\prime}(R)+\frac{m}{6(1+m)} R f^{\prime}(R)-\frac{f(R)}{6}=-\frac{\rho}{3}
\end{equation}

Now, if we assume the Energy-momentum tensor corresponding to the fluid distribution is conserved independently, 
i.e., the equation $\dot{\rho} + 3H (\rho + p)=0$ is satisfied, then for a dust dominated (pressure $p=0$) case, 
$\rho=\dfrac{\rho_0}{a^3}$ and equation \eqref{Rceq3} takes the form,
\begin{equation}\label{const-acc-fdprime}
 R^2f^{\prime\prime}(R)+\frac{m}{2(1-m)} R f^{\prime}(R)-\frac{1+m}{2(1-m)}f(R)=ER^{\frac{3}{2(1-m)}},
\end{equation}
where $E=-\dfrac{\rho_0}{C^3}\dfrac{(1+m)}{(1-m)}
\left[\dfrac{(1-m)^2}{6(1+m)}\right]^\frac{3}{2(1-m)}.$ \\

The equation (\ref{const-acc-fdprime}) can be integrated analytically and the solution for $f(R)$ is given by,
\begin{equation}\label{solconacc}
f(R)=C_1 R^\alpha+C_2 R^\beta+F R^\gamma,
\end{equation}
where $C_1$, $C_2$ are integration constants.  
Constants $F, \alpha , \beta , \gamma$ are given by,\\[2ex]
$F=\dfrac{4E(1-m)^2}{1+m(9+2m)}$,\\[2ex]
$\alpha=\dfrac{1}{4(1-m)}\left[(2-3m)+\sqrt{m^2-12m+12}\right]$\\[2ex]
$\beta=\dfrac{1}{4(1-m)}\left[(2-3m)-\sqrt{m^2-12m+12}\right]$\\[2ex]
$\gamma=\dfrac{3}{2(1-m)}$.\\

\begin{figure}
  \includegraphics[height=6cm, width=1.04\columnwidth]{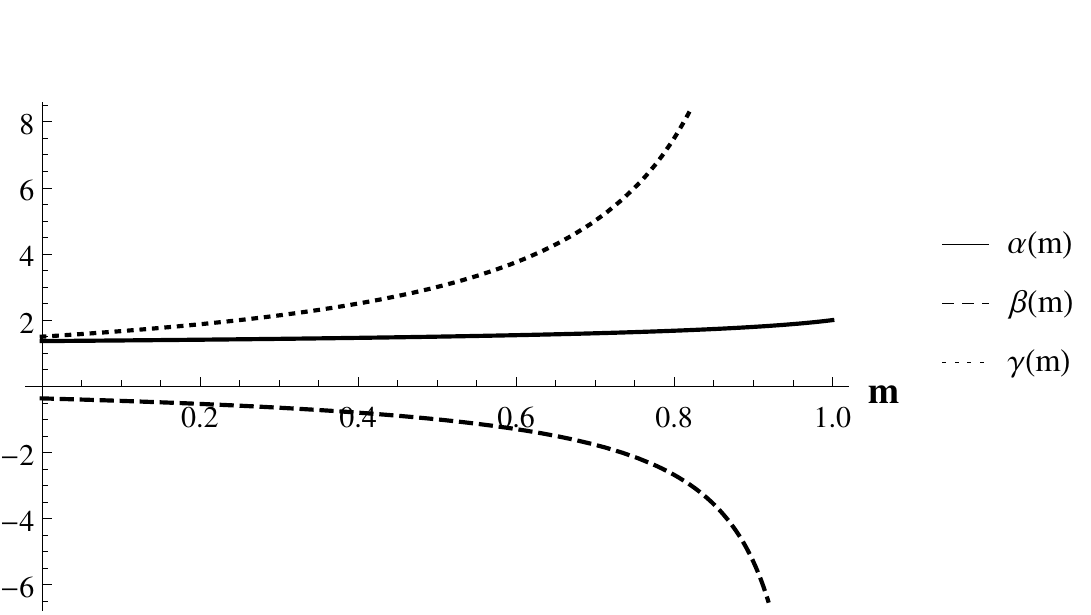}
 \caption{Plot of $\alpha$, $\beta$, $\gamma$ with $m$}
 \label{fig1}
\end{figure}

The solution for $f(R)$ contains three different powers of $R$. Figure \ref{fig1} shows the variation of $\alpha$, $\beta$ 
and $\gamma$ with $m$.
It is to be  noted that for $0<m<1$, $\alpha$ and $\gamma$ are 
always positive, while $\beta$ is always negative. It may be pointed out that  $C_2$, being a constant of integration, 
can be  chosen to be  zero, so we find that it is possible to have  accelerated expansion even with an action that 
contains only positive powers of $R$. 
It deserves mention that \citet{capozz} already observed this. \\

In order to have General Relativity as a special case from this particular class of models,  one of the two positive 
powers must be unity. 
From the plots, we find that $\alpha$ can not be unity in the given range of $m$. If one wants to have $\gamma = 1$, 
$m$ turns out to be $-\dfrac{1}{2}$, and the model would not yield an accelerated expansion. \\

Here it should be mentioned that we can always take $m = 1$ in equation (\ref{q}), that will give us an
exponential expansion, a pure deSitter universe without any matter that evolves with time, which is not included in the discussion.\\

{ \it Viability analysis:}\\

{ A model with a constant acceleration is definitely not  one that the observations indicate. We shall try to check 
if the corresponding $f(R)$ gravity model is  theoretically  consistent. For any $f(R)$ model 
to be viable one must have $f'(R)>0$ and $f''(R)>0$ \citep{hu, silv, faraoni}. The first condition ensures the effective 
constant of gravitation is positive and second condition is needed for the stability of the model. 
In the expression of $f(R)$ \eqref{solconacc}, the second term
will dominate as $R\rightarrow 0$ as $\beta$ is negative. Thus for the viability criterion at low curvature we 
must have $C_2=0$ and $C_1>0$.
whereas at high curvature the third term will dominate as $\gamma>\alpha$ always. But the coefficient  $F<0$, which means the
model will not be viable at high curvature which is also illustrated in figures \ref{sv1} and \ref{sv2} where for example we
have chosen $m=0.5$ and $\dfrac{\rho_0}{C^3}=1$ for the plots.
\begin{figure}
  \includegraphics[height=5.5cm,width=1.04\columnwidth]{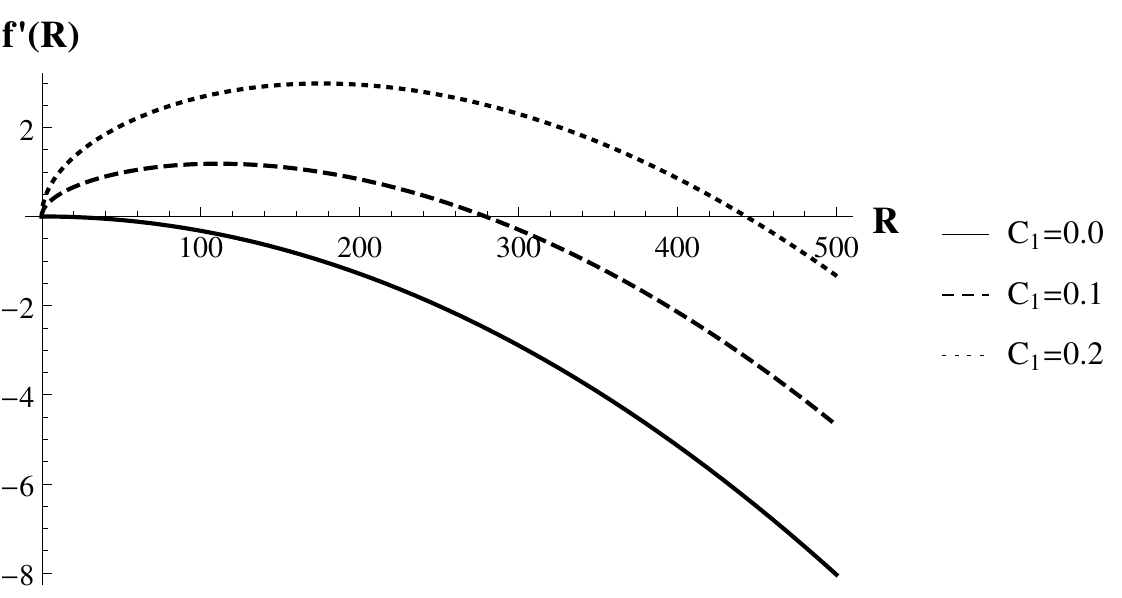}
 \caption{{Plot of $f'$ with $R$ for different values of $C_1$, where we have chosen $C_2=0$, $m=0.5$ and $\dfrac{\rho_0}{C^3}=1$.}}  
 \label{sv1}
\end{figure}
 \begin{figure}
 \includegraphics[height=5.5cm,width=1.04\columnwidth]{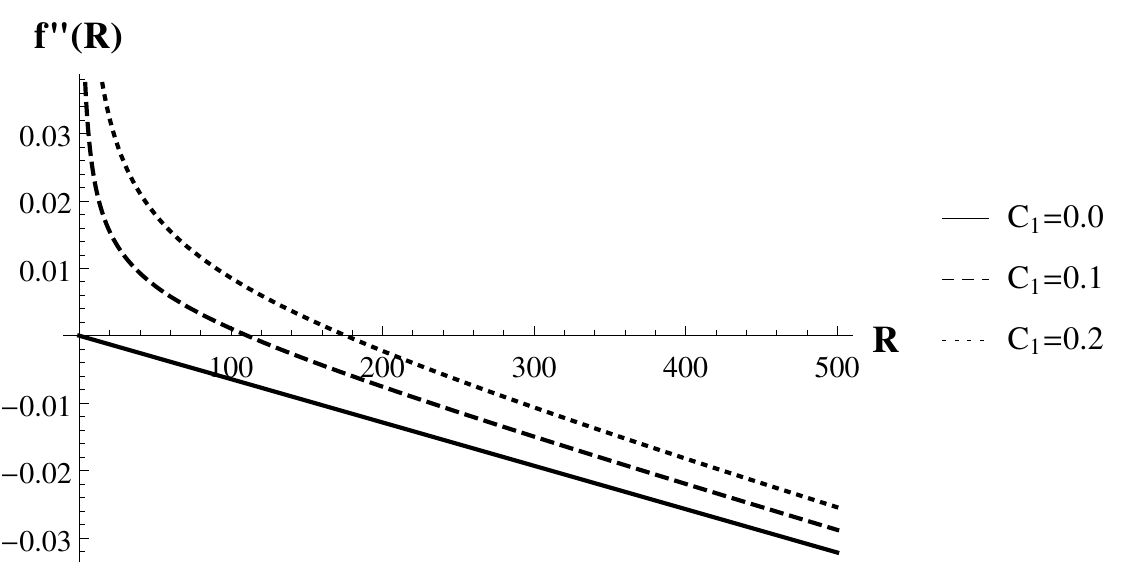}
 \caption{{Plot of $f''$ with $R$ for different values of $C_1$, where we have chosen $C_2=0$, $m=0.5$ and $\dfrac{\rho_0}{C^3}=1$.}} 
 \label{sv2}
\end{figure}}

\subsection{A constant jerk parameter}\label{jerk}
We have assumed $q$ to be constant in the previous section to find the relevant $f(R)$.
 Now that $q$ can be estimated from the observational data and is found to be evolving, the focus should 
 naturally shift to its evolution, namely the third order derivative of $a$, 
given by the dimensionless jerk parameter $j$ as,
\begin{equation}\label{const_j}
 j=\frac{1}{H^3}\frac{\dddot{a}}{a}.
\end{equation}

The jerk parameter finds increasing interest in the reconstruction of the models of the universe with 
an accelerated expansion. We refer to the references \citet{zhang} and \citet{ankan} for the motivation behind 
treating $j$ as an important kinematical quantity as the starting block for the reconstruction of models with an accelerated expansion. \\

It is well known that in spite of the huge discrepancy between the theoretically predicted value and the cosmologically 
required one of the cosmological constant $\Lambda$, a $\Lambda$CDM model does very well in 
explaining the accelerated expansion of the universe. In what follows, we shall assume 
\begin{equation}\label{j1}
 j=\frac{1}{H^3}\frac{\dddot{a}}{a}=1
\end{equation}
which mimics the
$\Lambda$CDM model, and make an attempt to reconstruct the corresponding $f(R)$ gravity model. \\

The general solution of \eqref{j1} is, 
\begin{equation}\label{scale}
 a(t)=\left[A\exp(\lambda t)+B\exp(-\lambda t)\right]^{\frac{2}{3}},
\end{equation}
where $A, B,$ and  $\lambda$ are integration constants. We note in passing that if $AB<0$  
we can rewrite this expression in the form $a(t) = a_0 \left[\sinh(\lambda(t-t_0))\right]^{\frac{2}{3}}$, we shall call this as 
Type I evolution, while 
for $AB>0$  
we can write $a(t) = a_0 \left[\cosh(\lambda(t-t_0))\right]^{\frac{2}{3}}$, we shall call this
as Type II evolution of the scale factor. \\

{ We can calculate the effective equation of state parameter using the expression for the scale factor \eqref{scale} 
in the same way as in the previous case which is given by,
\begin{equation}
 w_\mathrm{eff} = \frac{p_\mathrm{eff}}{\rho_\mathrm{eff}}=
 -\frac{\left[A\exp(\lambda t)+B\exp(-\lambda t)\right]^2}{\left[A\exp(\lambda t)-B\exp(-\lambda t)\right]^2}
\end{equation}
For Type I evolution we have,
\begin{equation}
 \frac{p_\mathrm{eff}}{\rho_\mathrm{eff}}=
 -\tanh[\lambda(t-t_0)]^2,
\end{equation} 
which tends to $-1$ when $t\rightarrow\infty$ and tends to zero when $t\rightarrow t_0$. This looks quite promising as we have a 
long matter era followed by an accelerated expansion, which is expected from a $\Lambda$CDM model. In this case,
\begin{equation}\label{encond1}
 \rho_\mathrm{eff}+3p_\mathrm{eff}=\frac{4}{3} \lambda ^2 \left(\coth ^2[\lambda  (t-{t_0})]-3\right)
\end{equation}}

{ For Type II evolution,
\begin{equation}
 w_\mathrm{eff} = \frac{p_\mathrm{eff}}{\rho_\mathrm{eff}}=
 -\coth[\lambda(t-t_0)]^2
 \end{equation}
which also tends to $-1$ when $t\rightarrow\infty$ but tends to a very large value as $t\rightarrow t_0$. 
This is definitely unacceptable as a model for the observed universe, as one does not have a matter dominated era in the past. Here we have,
\begin{equation}\label{encond2}
 \rho_\mathrm{eff}+3p_\mathrm{eff}=\frac{4}{3} \lambda ^2 \left(\tanh ^2[\lambda  (t-{t_0})]-3\right)
\end{equation}
For Type I evolution, equation (\ref{encond1}) indicates that the energy condition is satisfied or violated depending on the epoch $t$ one is looking at. Whereas for Type II evolution, the energy condition will always be violated as can be seen from equation (\ref{encond2}). \\}

The Ricci scalar for Type I evolution is then,
\begin{equation}\label{ricciI}
 R=\frac{16\lambda^2}{3}\left(1+\frac{a_0^3}{a^3}\right).
\end{equation}
and for Type II evolution,
\begin{equation}\label{ricciII}
 R=\frac{16\lambda^2}{3}\left(1-\frac{a_0^3}{a^3}\right).
\end{equation}
As the scale factor has to be real and positive, the following conditions have to be satisfied :  
for Type I evolution $R>\dfrac{16\lambda^2}{3}$ while for Type II evolution $R<\dfrac{16\lambda^2}{3}$.  \\

For both the cases the Raychaudhuri equation \eqref{rceq1} takes the form,
\begin{equation}\label{mastereq}
\begin{split}
 \left(R-{4}\lambda^2\right)\left(R-\frac{16}{3}\lambda^2\right)f^{\prime\prime}(R)-\frac{1}{6}\left(R-8\lambda^2\right) f^\prime(R)\\
 -\frac{1}{6}f(R)=-\frac{\rho}{3}.
 \end{split}
\end{equation}
The corresponding homogeneous equation (i.e, $\rho=0$ in this case) can be transformed into the standard hypergeometric
equation, by the substitution  $z=\dfrac{3(R-4\lambda^2)}{4\lambda^2}$, as
\begin{equation}\label{mastereqz}
 z(1-z)\frac{\mathrm{d}^2 f}{\mathrm{d}z^2}+[c-(a+b+1)z]\frac{\mathrm{d}f}{\mathrm{d}z}-ab f=0 ,
 \end{equation}
where $a=\dfrac{-7+\sqrt{73}}{12}$, $b=\dfrac{-7-\sqrt{73}}{12}$,
$c=-\dfrac{1}{2}$.

If the argument $z$ is complex, this hypergeometric equation has three different singular points at $z=0, 1, \infty$. 
In terms of $R$, they are at $R=4\lambda^2, \dfrac{16}{3}\lambda^2, \infty$. But here we are interested in real solutions, thus 
one has to distinguish between $\pm \infty$ and the homogeneous part of equation \eqref{mastereq} has real solutions around 
four different singular points, namely, $R=-\infty, 4\lambda^2, \dfrac{16}{3}\lambda^2, \infty$. 
The solutions around different singular points and their region of validity
are summarised in TABLE I. For a discussion on hypergeometric functions, 
we refer to the work of \citet{maier}. \\ \\

\begin{table*}
\label{tab}
\caption{The solution around different singular points and their region of validity. Subscript $h$ stands for the homogeneous part
of the solution. $W_1$ and $W_2$ are integration constants.}
 \centering
 \begin{tabular}{|c|c|c|}
\hline 
Singular Point & Solution & Range of applicability  \\ 
\hline 
$z=0$ $\left(R=4\lambda^2\right)$ &  $f_{h}(z)=W_1\hspace{0.2cm} {}_2F_1\left( a,b; c; z \right)
 $ & $-1<z<1$ $\left(\dfrac{8\lambda^2}{3}<R<\dfrac{16\lambda^2}{3}\right)$ \\ 
 & $+W_2 z^{1-c} {}_2F_1\left(1+a-c,1+b-c; 2-c; z \right)$ & \\[2ex]
\hline 
$z=1$ $\left(R=\dfrac{16\lambda^2}{3}\right)$ &  $f_{h}(z)=W_1\hspace{0.2cm} {}_2F_1\left( a,b; 1+a+b-c; 1-z \right)
 $ & $0<z<2$ $\left(4\lambda^2<R<\dfrac{20\lambda^2}{3}\right)$ \\ 
 & $+W_2 \left(z-1\right)^{c-a-b} {}_2F_1\left(c-a,c-b; 1+c-a-b; 1-z \right).$ &  \\[2ex]
\hline 
$z=\infty$ $\left(R=\infty\right)$ &  $f_{h}(z)=W_1 \left(z-1\right)^{-a}
 {}_2F_1\left( a,c-b; a-b+1; \left(1-z\right)^{-1}  \right)
 $ & $1<z<\infty$ $\left(\dfrac{16\lambda^2}{3}<R<\infty\right)$ \\ 
 & $+W_2 \left(z-1\right)^{-b}
 {}_2F_1\left( b,c-a; b-a+1; \left(1-z\right)^{-1}  \right)$ &   \\ [2ex]
\hline 
$z=-\infty$ $\left(R=-\infty\right)$ &  $f_{h}(z)=W_1 (-z)^{-a}
 {}_2F_1\left( a,a-c+1; a-b+1; {z}^{-1}  \right)
 $ & $-\infty<z<0$ $\left(-\infty<R<4\lambda^2\right)$ \\ 
 & $+W_2 (-z)^{-b}
 {}_2F_1\left( b,b-c+1; b-a+1; {z}^{-1}  \right)$ & \\[2ex]
\hline
\end{tabular} 
\end{table*}

Here we have written down four different solutions around four singular points. Now, the question that which of these 
solutions 
are actually relevant as the complementary functions of equation \eqref{mastereq} depends on the boundary conditions and what 
range 
of the Ricci scalar $R$ one is looking for.  We will discuss this in detail when we write down the general solution 
for $f(R)$.\\

We will now solve for the particular integral by considering a particular form for 
the inhomogeneous term $\dfrac{\rho}{3}$ in the right hand side of equation \eqref{mastereq}. Here again
we consider $\rho=\dfrac{\rho_0}{a^3}$ as discussed in the context of constant deceleration parameter case and equation \eqref{mastereq} 
becomes,
\begin{equation}
\label{rceq-j}
 \begin{split}
 \left(R-{4}\lambda^2\right)\left(R-\frac{16}{3}\lambda^2\right)f^{\prime\prime}(R)-\frac{1}{6}\left(R-8\lambda^2\right)
 f^\prime(R)\\-\frac{1}{6}f(R)= k \left(R-\frac{16\lambda^2}{3}\right)
 \end{split}
\end{equation}
where $ k =\pm\dfrac{\rho_0}{16\lambda^2 a_0^3}$, positive and negative sign correspond to Type II and Type I evolution
respectively. The particular integral
for this equation can easily be found to be,
\begin{equation}
 f_{p}(R)=-3k\left(R-\frac{8\lambda^2}{3}\right)
\end{equation}
where subscript $p$ stands for particular integral. \\

In order to find the relevant $f(R)$ gravity model giving rise to a late time $\Lambda$CDM model, we need to choose proper 
conditions.
We will illustrate this with two examples, one each for Type I and Type II evolution. \\

{\bf Example of a general solution for Type I evolution:}

Let us first take up the Type I evolution. From Table I, we have two choices for this case.
We can use the third solution for the whole 
range $\dfrac{16\lambda^2}{3}<R<\infty$ and also the second solution for the range 
$\dfrac{16\lambda^2}{3}<R<\dfrac{20\lambda^2}{3}$.
We will use the third solution as the complementary function for this case, as this 
one function will do the job for the entire
region $R>\dfrac{16\lambda^2}{3}$.\\

We will thus write down the complete solution for $f(R)$, in terms of $z=\dfrac{3(R-4\lambda^2)}{4\lambda^2}$,
for Type I case which is valid for the whole range $R>\dfrac{16\lambda^2}{3}$. As the curvature $R$ is expected to decrease 
with the evolution, this is consistent with an indefinite past -
\begin{equation}\label{finalfr}
 \begin{split}
  f(z)=W_1 \left(z-1\right)^{-a}
 {}_2F_1\left( a,c-b; a-b+1; \left(1-z\right)^{-1}  \right)\\+W_2 \left(z-1\right)^{-b}
 {}_2F_1\left( b,c-a; b-a+1; \left(1-z\right)^{-1}  \right)\\-4k\lambda^2\left(z+1\right).
 \end{split}
\end{equation}

We have one free parameter at our disposal to make this work at the present epoch. 
The value of $\lambda$ should be such that that $R_0 > \dfrac{16}{3}\lambda^2$, where $R_0$ is the present value of the Ricci scalar. \\

{\it Viability of the solution:}\\

{ From the expression of $f(z)$, one can note that at high curvature when $z\rightarrow \infty$ 
the second term will dominate. Thus, for the viability criterion at high curvature regime we must have $W_2>0$. 
At low curvature when $z$ is very close to $1$, the first two terms will
dominate and the behaviour will depend on the relative sizes of $W_1$ and $W_2$. In this case $f'$ and $f''$ will be of opposite 
signs as can be seen from the figures \ref{sv3}, \ref{sv4} where as example we have chosen $W_2=0$ and $\dfrac{\rho_0}{a_0^3}=1$
(with this choice $\dfrac{\rho_0}{a_0^3}=1$ we have $f'_p(z)=1$ but the qualitative inferences will not depend on this particular choice)
and thus  both of them can not be positive simultaneously. So either $f'$ or$f''$ dips to high
negative values for low values of $z$.
The model will show an effectively negative value of the Newtonian constant of gravity or will be unstable, 
which are not viable options for local astronomy if not for anything else.
\begin{figure}
  \includegraphics[height=5.5cm,width=1.04\columnwidth]{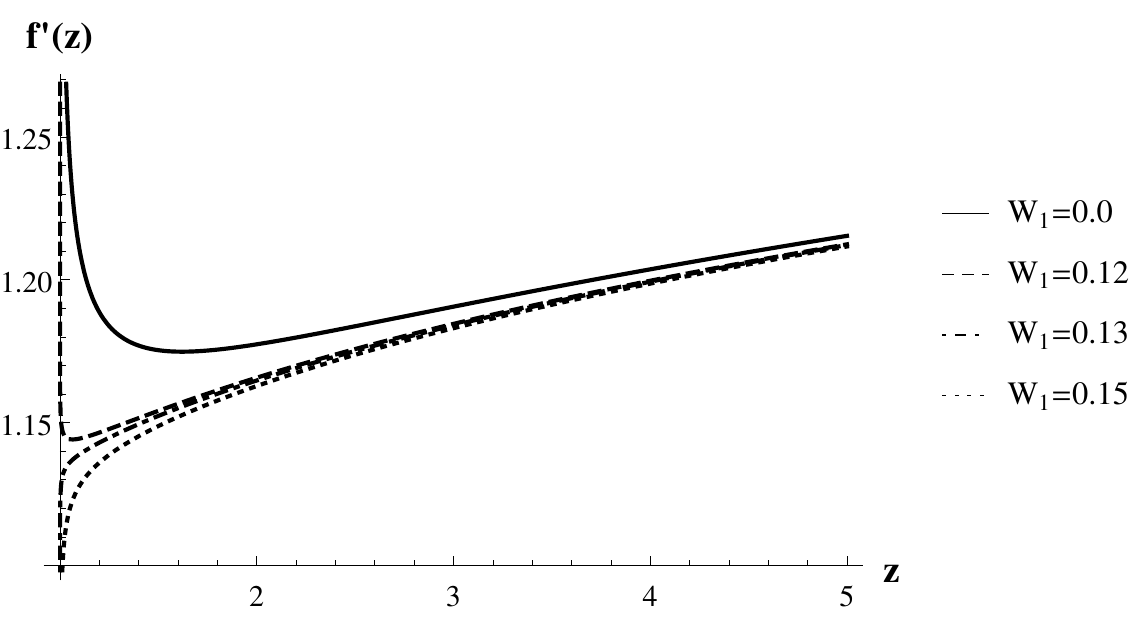}
 \caption{{Plot of $f'$ with $z$ for different values of $W_1$, where we have chosen $W_2=0.1$, and $4k\lambda^2=-\dfrac{\rho_0}{a_0^3}=-1$.}}  
 \label{sv3}
\end{figure}
 \begin{figure}
 \includegraphics[height=5.5cm,width=1.04\columnwidth]{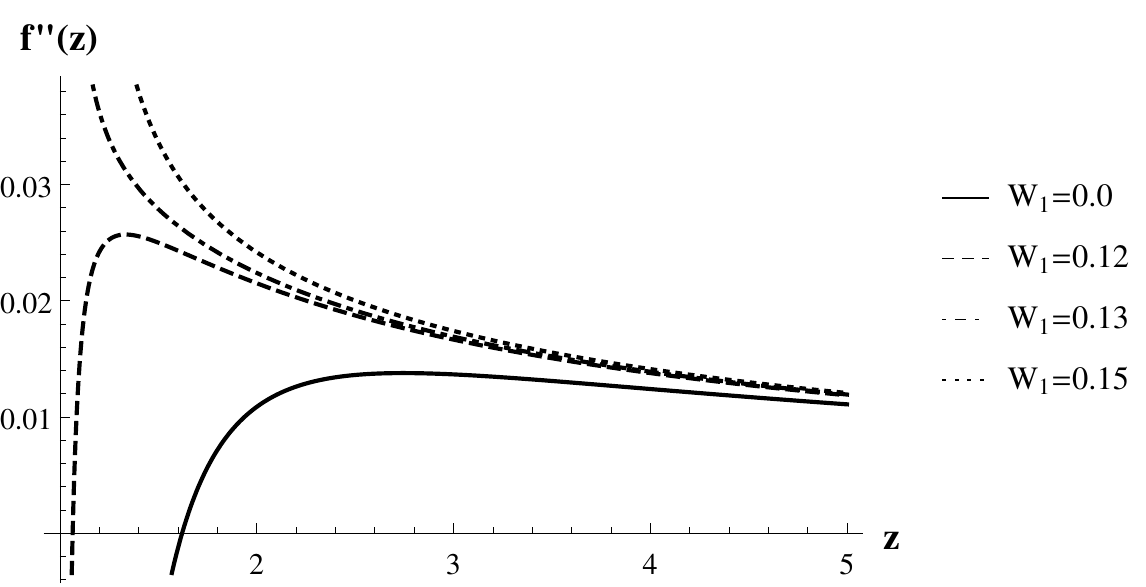}
 \caption{{Plot of $f''$ with $z$ for different values of $W_1$, where we have chosen $W_2=0.1$, and $4k\lambda^2=-\dfrac{\rho_0}{a_0^3}=-1$.}} 
 \label{sv4}
\end{figure}\\}

{\bf Example of a general solution for Type II evolution:}

 For Type II case, we can use  the
first solution for the range $\dfrac{8\lambda^2}{3}<R<\dfrac{16\lambda^2}{3}$, the second solution
for the range $4\lambda^2<R<\dfrac{16\lambda^2}{3}$ and the fourth solution for the range $-\infty<R<4\lambda^2$.
None of which are valid for the whole range $0<R<\dfrac{16\lambda^2}{3}$.
Here we will use the first one as the complementary function, 
so the complete solution which is valid for the range $\dfrac{8\lambda^2}{3}<R<\dfrac{16\lambda^2}{3}$,
is given by,
\begin{equation}\label{finalfr2}
 \begin{split}
  f(z)=W_1\hspace{0.2cm} {}_2F_1\left( a,b; c; z \right)
 \\+W_2 z^{1-c} {}_2F_1\left(1+a-c,1+b-c; 2-c; z \right)\\-4k\lambda^2\left(z+1\right).
 \end{split}
\end{equation}
{ For $W_2 \neq 0$ this expression is valid in the range $0<z<1$. Otherwise it is valid for the range $-1<z<1$.}

One can easily see that by fixing the value of $\lambda$, this example can also be made work at the present epoch, 
but this cannot be extended to an indefinite past.\\

{ \it Viability of the solution:}\\

{ With non-zero $W_2$, when $z\rightarrow 0$, the contribution from the second term in $f'$ will be very small. 
If we choose, for example 
$4k\lambda^2=\dfrac{\rho_0}{a_0^3}=1$,
some manipulations with the hypergeometric series will reveal that $W_1>3$ for $f'>0$ when $z\rightarrow 0$. The contribution
from the second term will dominate in $f''$. Thus, we need $W_2>0$ for $f''$ to be positive. When $W_2=0$ for $f''>0$
we must have $W_1<0$. Thus we cannot make both $f'$ and $f''$ positive when $z\rightarrow 0$ with $W_2=0$. We have also plotted
$f'$ and $f''$ in figures \ref{sv5} and \ref{sv6} to study the features in more detail with the choice $4k\lambda^2=\dfrac{\rho_0}{a_0^3}=1$
and $W_1=3.5$ as example. For higher values of $W_2$, such as $2$ or $3$, 
one can have both $f'$ and $f''$ positive, but the equation of state for the type II models are completely unfavourable for low values 
of $t$, i.e., high values of $z$.
\begin{figure}
  \includegraphics[height=5.5cm,width=1.04\columnwidth]{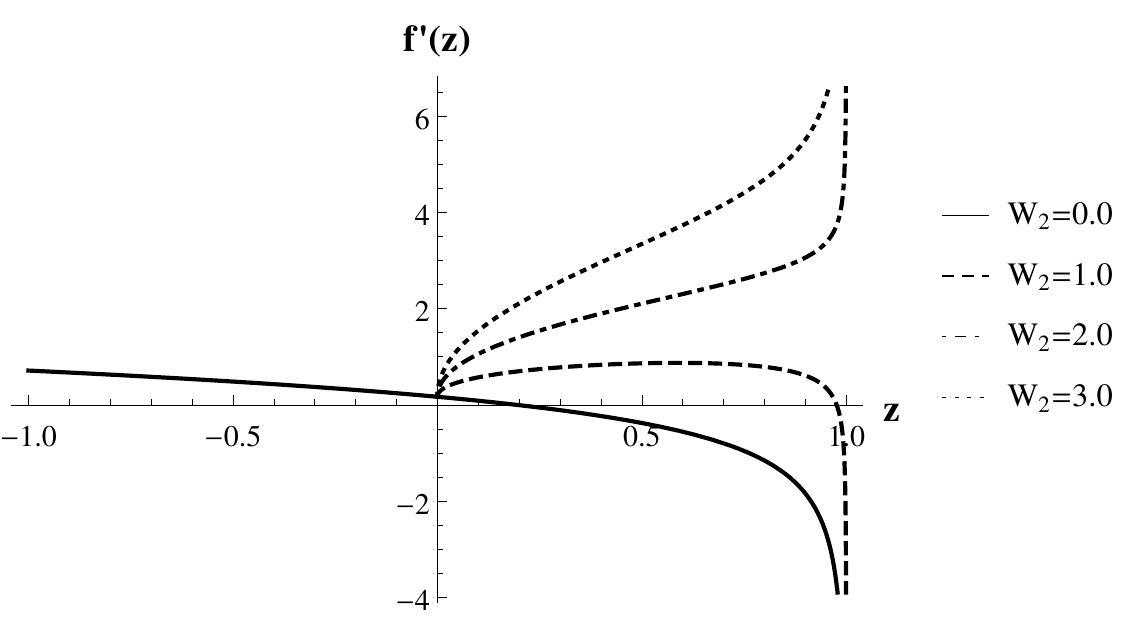}
 \caption{{Plot of $f'$ with $z$ for different values of $W_2$, where we have chosen $W_1=3.5$, and $4k\lambda^2=\dfrac{\rho_0}{a_0^3}=1$.}}  
 \label{sv5}
\end{figure}
 \begin{figure}
 \includegraphics[height=5.5cm,width=1.04\columnwidth]{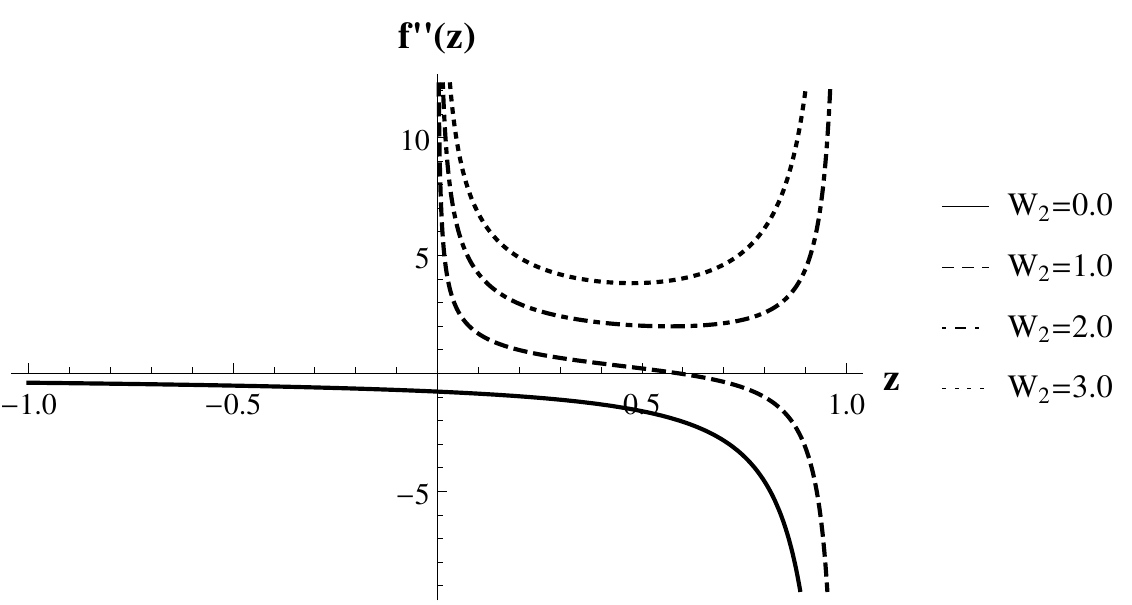}
 \caption{{Plot of $f''$ with $z$ for different values of $W_2$, where we have chosen $W_1=3.5$, and $4k\lambda^2=\dfrac{\rho_0}{a_0^3}=1$.}} 
 \label{sv6}
\end{figure}\\}

{ As examples we have chosen two of the solutions from Table I and performed the viability analysis.
We can also do the same for rest of the solutions of the table. For the solution around $z=1$ i.e. the second one in the table, 
$f'$ and $f''$ have opposite sign as $z\rightarrow 1$. The same is the situation for the solution
around $z=-\infty$ i.e. the last solution in the table, when $z\rightarrow 0$.\\}

One thing is important to note here that the particular integral for the constant
jerk case contains $R$ as a term, thus we can always recover the General Relativity
$f(R)=R$ as a special case for this kind of model.\\

\citet{Dunsby} started with an evolution ansatz  for the Hubble parameter. From Friedmann equations 
for an $f(R)$ gravity model along with a cosmological constant, they found that the only real valued $f(R)$ that is able to 
mimic $\Lambda$CDM expansion for a dust filled universe actually corresponds to the Einstein-Hilbert action with a positive 
cosmological constant. In a later work \citet{He} showed, with a slightly different approach, 
that there is indeed a 
real-valued analytical $f(R)$ in terms of the hypergeometric functions which admit an exact $\Lambda$CDM expansion history. 
The solution that they got matches with one of the present cases, written as equation \eqref{finalfr}, if one identifies 
$\dfrac{16}{3}\lambda^2=4\Lambda$ and $3k=-1$. The difference is that, they have made $W_2=0$ by arguing that the model 
should have a ``chameleon'' property i.e. $f_h(R)$ and $f_h^\prime(R)$ are convergent when $R\rightarrow \infty$.

\section{\label{sec:level4} Conclusion}
In this work a new strategy for reconstructing $f(R)$ gravity models for a given expansion history of the universe has been 
attempted 
using Raychaudhuri equation. Two examples have been successfully worked out. The first one is that of a simple ever 
accelerating model. 
The reconstructed $f(R)$ is a simple combination of powers of $R$, consistent with the examples found in the literature.
But potentially
a wide variety of models can be found from the present work as the powers of $R$ are not uniquely determined. 
One intriguing feature is 
that the $f(R)$ models, giving rise to Einstein gravity as a special case, cannot yield an ever accelerating model for 
the universe. \\ 

The second example recovers the celebrated $\Lambda$CDM mode of evolution. Two illustrations are given. The $f(R)$ theory 
recovered is 
definitely not a simple function of $R$, all the models are such that $f(R)$ is a hypergeometric function of $R$. We also 
recover the 
model given by \citet{He} as a special case of the first illustration (Type I). The model is valid for the early 
universe (large 
curvature regime) to the present epoch subject to a tuning of the constants. The second illustration  (Type II) is definitely
different 
from the one given by \citet{He}. But this works only for a limited span of the evolution as $R$ is bounded on both sides.
This looks fine for the current state of the universe, but cannot be extended to a distant past. \\

{ The major conclusion from the present work is the following. The second example that we discuss, 
which is arguably the most favoured 
evolution history of the universe, namely the $\Lambda$CDM mode of evolution, can lead only to the trivial choice $f(R)=R-2\Lambda$ as the viable option.
All the non trivial possibilities of the choice of $f(R)$ leading to $j=1$, are plagued with either instability, or a 
negative effective Newtonian constant of gravity or not having a sufficient matter dominated regime in the past or some 
combination of such pathologies. Our first example,  the toy model with a constant negative deceleration parameter, fails the fitness test of 
stability for moderately high values of $z$.\\}

The method, a theoretical reconstruction of cosmological models using the Raychaudhuri equation, appears to be quite powerful. 
Many exotic models can in principle be put to test with the help of this tool.

\section*{Acknowledgements} The authors thank the anonymous referee whose comments and suggestions improved the quality of the work. 
Shibendu Gupta Choudhury (SGC) thanks CSIR, India for financial support. SGC thanks Soumya Chakrabarti for valuable discussions.

\label{lastpage}
\end{document}